\documentclass{appolb}
\usepackage{graphicx}

\usepackage{float}
\usepackage{subfig}
\usepackage{authblk}
\usepackage{amsmath,bm}

\newcommand{\rf}[1]{(\ref{#1})}
\newcommand{\beq}{\begin{equation}}
\newcommand{\beql}[1]{\beq\label{#1}}
\newcommand{\eeq}{\end{equation}}
\newcommand{\bea}{\begin{eqnarray}}
\newcommand{\eea}{\end{eqnarray}}

\newcommand{\cD}{{\cal D}}

\newcommand{\cM}{{\cal M}}

\newcommand{\cT}{{\cal T}}

\newcommand{\cZ}{{\cal Z}}

\begin{document}
\title{Phase structure of Causal Dynamical Triangulations  in 4D%
\thanks{Presented at the 3rd conference of the Polish Society on Relativity}%
}
\author{Jakub Gizbert-Studnicki
\address{The M. Smoluchowski Institute of Physics, Jagiellonian University
\newline
 ul. prof. Stanis\l awa \L ojasiewicza 11, Krak\'ow, PL 30-348
 \newline
 Email: jakub.gizbert-studnicki@uj.edu.pl}
}
\maketitle
\begin{abstract}
Causal Dynamical Triangulations (CDT) is a lattice approach to quantum gravity. CDT has   rich phase structure, including a semiclassical phase consistent with Einstein's general relativity. Some of the observed phase transitions are second (or higher) order which opens a possibility of investigating the ultraviolet continuum limit.   Recently a new phase with intriguing geometric properties has been discovered and the new phase transition is also second (or higher) order.
\end{abstract}
  
\section{Introduction}

Quantum field theory (QFT) techniques provide powerful tools in describing three out of four fundamental interactions. The key problem in applying these methods to quantize gravity is that QFT based on Einstein's general relativity (GR) is perturbatively nonrenormalizable \cite{'tHooft:1974bx}. However, following Weinberg's asymptotic safety conjecture  \cite{Weinberg79}, it is possible that in the space of gravitational couplings there exist  non-Gaussian  ultraviolet (UV) fixed point(s) at which nonperturbative QFT techniques  could be used  to define a quantum  theory of gravity valid for any energy scale.\footnote{There are known examples of perturbatively nonrenormalizable but asymptoticaly safe  QFTs  and  evidence  is growing that it is also the case for gravity  \cite{Reuter06}.  } Among such techniques lattice methods play an increasingly important  role.
A good test of any lattice  approach is its ability to 
 reproduce  GR in the infrared limit  and also the existence 
 of second (or higher) order phase transitions  associated with the perspective UV fixed point(s).\footnote{Infinite correlation lengths characteristic of such phase transitions make it in principle possible to decrease lattice spacing  to zero, i.e. to investigate the continuum limit.} 
Therefore a study of the phase structure and the order of phase  transitions constitute  first steps in the quest for the continuum theory of quantum gravity. 
\section{Causal Dynamical Triangulations}

Causal Dynamical Triangulations (CDT) is a lattice model  based on the path integral quantization applied to GR. CDT gives a precise meaning to  a (formal) gravitational path integral
\beql{ZCDT}
\cZ_{GR}= \int \cD_{\cM}{[g]} e^{iS_{HE[g]}}\quad \rightarrow \quad \cZ_{CDT} = \sum_{\cT}e^{iS_R[\cT]} \ ,
\eeq
approximating continuous  geometries (described by all physically distinct metric tensors $g$) by  lattices constructed from two types of identical four-dimensional simplicial blocks
glued together to form triangulations $\cT$.
The key assumption of CDT is an introduction of causal structure by foliating spacetime into Cauchy hyper-surfaces $\Sigma$ of constant global proper time $T$.  Topologically a triangulation $\cT$ is $\Sigma \times T$, and one  requires that the topology of each spatial slice $\Sigma$ is fixed. 
 Each spatial layer $\Sigma$ at integer (lattice) time $t$ is by definition constructed from equilateral tetrahedra. The four-simplices interpolate between consecutive spatial  layers of integer $t$  in such a way that also all intermediate Cauchy layers between $t$ and $t+1$ have the requested fixed spatial topology. This can be done by using just two types of four-simplices called the $(4,1)$ simplex and  the $(3,2)$ simplex.\footnote{The numbers in parentheses  denote vertices lying in $t$ and $t\pm 1$, respectively.} It is assumed that the interior of each four-simplex is a flat Minkowski  spacetime and local curvature is defined by the way the simplices are glued together. In Eq. \rf{ZCDT} $S_R$ is the discretized Hilbert-Einstein action $S_{HE}$ obtained  following Regge's method for describing piecewise linear geometries \cite{Regge:1961px}
\beql{SRegge}
S_{R}=-\left(\kappa_{0}+6\Delta\right)N_{0}+\kappa_{4}\left(N_{(4,1)}+N_{(3,2)}\right)+\Delta \ N_{(4,1)} \  ,
\eeq 
where $ N_{(4,1)}$,  $ N_{(3,2)}$ and $N_0$ denote the total number of $(4,1)$ simplices, $(3,2)$ simplices and vertices, respectively, while
$\kappa_{0}$, $\Delta$ and $\kappa_4$ are three  bare coupling constants. They  are functions of the Newton's constant, the cosmological constant and the  asymmetry $\alpha$ between lengths of time-like and space-like links in the lattice ($a_t^2 = - \alpha\  a_s^2$). 

In order to study the regularised path integral one is forced to use  Monte Carlo techniques.
This can be done by  applying Wick rotation from positive to negative $\alpha$ values which changes time-like links
into space-like links, i.e. changes real (Lorentzian) time $t^{(L)}$ into imaginary (Euclidean) time $t^{(E)}$ ($t^{(L)} \to t^{(E)} = i t^{(L)}$), and also changes Lorentzian action into Euclidean action ($S_R^{(L)} \to S_R^{(E)} = i S_R^{(L)}$).
Accordingly, the path integral $\cZ_{CDT}$ \rf{ZCDT} becomes a partition function which can be studied numerically.

\begin{figure}[htb]
\centerline{
\includegraphics[width=7cm]{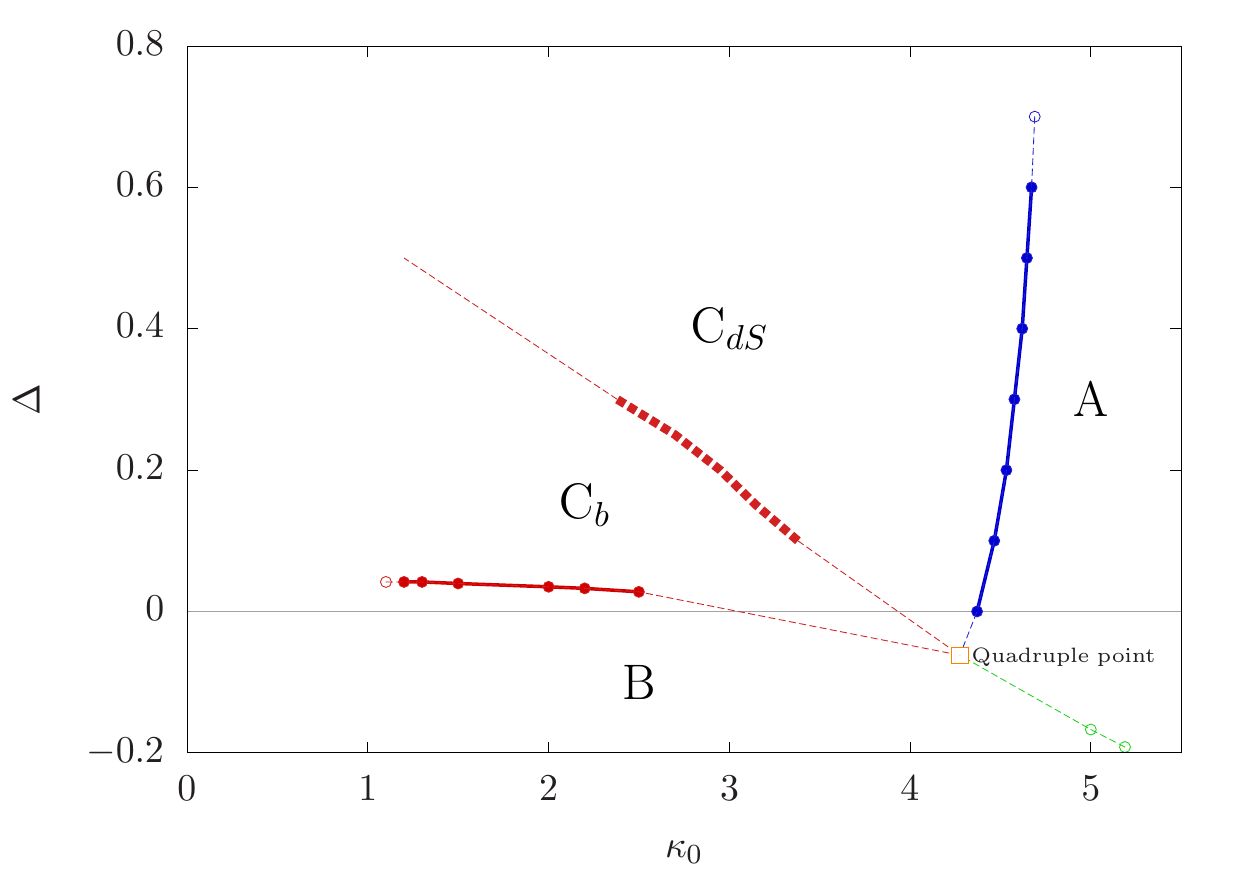}}
\caption{Phase structure of 4-dim CDT in the $(\kappa_0,\Delta)$ bare couplings plane.  $\kappa_4$ is fine-tuned to critical value consistent with infinite volume limit. }
\label{Fig:F1H}
\end{figure}

\section{Phase structure of 4-dim CDT}

All  results presented herein  were obtained for a particular choice of  fixed  spatial topology $\Sigma = S^3$ (3-sphere).\footnote{Preliminary studies of CDT with toroidal spatial topology confirm the existence of similar phase structure.} Historically, three phases of various spacetime geometry called $A$, $B$ and $C$  were discovered (see Fig. \ref{Fig:F1H}).
Phase $A$ (time uncorrelated geometry) and $B$ (time collapsed geometry) do not have clear physical interpretation. The most interesting one is phase $C$, which is separated from phase $A$ by a 1st order transition and from phase $B$ by a 2nd (or higher) order transition  \cite{Ambjorn:2012ij}. The basic  feature of phase $C$ (now $C_{dS}$) is the emergence of large scale four-dimensional geometry \cite{Ambjorn05} consistent with semiclassical (Euclidean) de Sitter universe \cite{Ambjorn:2008wc}. It was also shown that in this phase quantum fluctuation of spatial volume are governed by the minisuperspace reduction of the Hilbert-Einstein action \cite{Ambjorn:2008wc,Ambjorn:2011ph}. 

The key tool in the analysis of the effective action of CDT was the transfer matrix (TM) parametrised by a spatial volume observable, i.e. the transition amplitude from spatial volume $n$ at (lattice) time $t$ to spatial volume $m$ at time $t+1$. Deep inside phase $C$ ($C_{dS}$ region in Fig. \ref{Fig:F1H}) the TM  is well parametrised by \cite{JGSTM}:
\begin{equation}
\langle n|M_{C_{dS}}|m \rangle = \underbrace{\exp\Bigg[- \frac{1}{\Gamma} \frac{(n-m)^2}{ (n+m)}\Bigg] }_{\text{kinetic part}}  \underbrace{\exp\Bigg[ - \mu \left(\frac{n+m}{2}\right)^{1/3} + \lambda \left(\frac{n+m}{2}\right) \Bigg] }_{\text{potential part}} ,
\label{TMCdS}
\end{equation}
 where $\Gamma$, $\mu$ and $\lambda$ are parameters related to the Newton's constant, the size of the CDT universe and the cosmological constant, respectively. However it was observed that in some region of the parameter space close to phase $A$ ($C_{b}$ region in Fig. \ref{Fig:F1H}) the TM kinetic part bifurcates, such that \cite{Ambjorn:2014mra}:
\begin{equation}
\langle n|M_{C_b}|m \rangle =\left[ \exp \left( -\frac{1}{\Gamma}\frac{\Big((n-m) - \big[c_0 (n+m - s_b) \big]_+\Big)^2}{n+m}\right) + \right.
\label{TMCb}
\end{equation}
$$
+ \left. \exp \left( -\frac{1}{\Gamma}\frac{\Big((n-m) + \big[c_0 (n+m - s_b) \big]_+\Big)^2}{n+m}\right)\right]\times  \text{potential\_part}[n+m] \ ,
$$
which led to a discovery of a new phase transition associated with the parameters $s_b\to \infty$ and $c_0\to 0$, where TM (\ref{TMCb}) transforms into TM (\ref{TMCdS}). 

The newly discovered bifurcation phase $C_b$ has many intriguing geometric properties, including very large (potentially infinite)  Hausdorff dimension and also spectral dimension  becoming much larger than 4 for long diffusion times. The $C_{dS}-C_{b}$ phase transition is related to breaking of spatial homogeneity of phase $C_{dS}$ by the appearance  of  compact  spatial volume clusters concentrated around "singular" vertices with macroscopically large coordination number present  every  second time layer inside phase $C_b$ \cite{JHEP1508}.  The volume clusters have (topologically) spherical boundaries and thus can  technically  be called "black balls". The "black balls" around   singular vertices in time $t$ and $t+2$ are causally connected through those intermediate  4-simplices which also share   one of the "singular" vertices. As a result one observes a  marked-out four-dimensional geometric structure related to  evolution  of time-correlated "black ball" volume condensations. The geometry of  phase $C_b$ requires further   studies, but a working hypothesis is that what is  observed might be actually a  quantum black hole. 

A key question remains about the order of the recently discovered bifurcation transition. In Ref. \cite{Coumbe:2015oaa} an order parameter related to the appearance of high order vertices   was proposed:
\begin{equation}\label{op}
\text{OP}_2= \frac{1}{2} \left[ \Big| O_{max}\big(t_0\big) - O_{max}\big(t_0+1\big) \Big|   +     \Big| O_{max}\big(t_0\big) - O_{max}\big(t_0-1\big) \Big|\right],
\end{equation}
where $ O_{max}\big(t_0\big)$ is  the highest coordination number among all vertices and $O_{max}\big(t_0\pm 1\big)$ is the highest coordination number of a vertex observed in the neighbouring time slice. The position of the $C_{dS}-C_{b}$ transition is signalled by a peak in susceptibility $\chi_{\text{OP}_2} = \langle \text{OP}_2 \rangle - \langle \text{OP}_2 \rangle^2$ and it moves in the CDT bare couplings space $(\kappa_0,\Delta)$ when lattice volume $N_{(4,1)}$ is changed. The volume dependence of critical $\Delta$ (for $\kappa_0=2.2$ fixed) can be fitted with the following function: 
\begin{equation}\label{voldep}
\Delta^{crit}\big(N_{(4,1)}\big)=\Delta^{crit}\big(\infty\big) - \alpha \ N_{(4,1)}^{\ -1/\gamma} \ ,
\end{equation}
and  measured value of the critical exponent $\gamma = 2.71 \pm 0.34$ strongly supports the conjecture that the bifurcation transition is a 2nd (or higher) order phase transition.\footnote{For a 1st order transition one should have $\gamma=1$.}

\section{Summary and conclusions}
We have briefly presented the recently updated 4-dim CDT phase diagram, including the semiclassical phase $C_{dS}$ which seems to be  CDT realisation of the correspondence principle, and the newly discovered bifurcation phase $C_b$ with very nontrivial geometric properties. We have provided evidence that the $C_{dS}$-$C_{b}$ phase transition, related to breaking of homogeneity of semiclassical phase, is a 2nd (or higher) order phase transition which may in principle allow one to approach the perspective UV fixed point of quantum gravity \cite{Ambjorn:2016cpa}.

\section{Acknowledgements}
The author wishes to  acknowledge the support of the grant DEC- 2012/ 06/A/ST2/00389 from the National Science Centre Poland. The results described herein were obtained in close collaboration with J. Jurkiewicz, J. Ambj\o rn,  R. Loll,  A. G\"orlich and D.N. Coumbe.

\end{document}